\documentclass[amsmath,amssymb, aps,prapplied, twocolumn,superscriptaddress]{revtex4-2}
\usepackage[utf8]{inputenc}
\usepackage{hyperref}
\hypersetup{
	colorlinks=true,
	linkcolor=blue,
	filecolor=blue,
	citecolor = blue,      
	urlcolor=cyan,
}

\usepackage{amsmath}
\usepackage{amsfonts}
\usepackage{MnSymbol}
\usepackage{xcolor}
\usepackage{graphicx}

\newcommand{\Rb}{R_{b}}
\newcommand{\LAR}{\texttt{LearnAndReduce}~}

\newcommand{\figref}[1]{Fig.~\ref{#1}}

\newcommand{\secref}[1]{Sec.~\ref{#1}}

\newcommand{\jdp}[1]{{\color{black}{#1}}}
\newcommand{\jk}[1]{{\color{black}{#1}}}

\begin{document}

\title{Reducibility of native weighted graphs on Rydberg Arrays}

\author{J. Kombe}
\email{johannes.kombe@strath.ac.uk}
\affiliation{Department of Physics and SUPA, University of Strathclyde, Glasgow G4 0NG, UK}

\author{J. D. Pritchard}
\email{jonathan.pritchard@strath.ac.uk}
\affiliation{Department of Physics and SUPA, University of Strathclyde, Glasgow G4 0NG, UK}

\begin{abstract}
We investigate the classical reducibility of random unit-disk graph (UDG) instances of the maximum independent set (MIS) and maximum weighted independent set (MWIS) problems, which can be natively realised in Rydberg atom quantum processors. Using state-of-the-art kernelisation techniques, we systematically probe how far classical preprocessing can simplify such native optimisation problems of varying size and connectivity. While many small or sparse instances can be fully reduced, dense graphs often retain finite irreducible kernels even after extensive reductions. Introducing vertex weights tends to increase reducibility, whereas extending the interaction range in the underlying UDG connectivity suppresses the reduction efficiency. By exploring where classical reductions cease to be effective, we aim to delineate the regime of problem instances that remain computationally demanding—those most relevant for testing and benchmarking near-term quantum optimisation hardware. We find that for the remaining finite kernels, quantum execution would require non-native embeddings with substantial resource overheads, suggesting that directly running native instances may be more practical than embedding a reduced kernel. 
\end{abstract}
\maketitle

\section{Introduction}
\label{sec:intro}
Combinatorial optimisation problems lie at the core of computer science, mathematics, and operations research, encompassing a vast range of applications from logistics and network design to scheduling, finance, and drug discovery \cite{Paschos2014}. Since Karp’s seminal compilation of 21 NP-complete problems \cite{Karp1972}, it has been known that many such problems are computationally intractable in the worst case, with solution times that scale exponentially with system size \cite{PapadimitriouSteiglitz1998,KorteVygen2018}. Despite decades of progress in heuristic and approximation algorithms, the intrinsic hardness of combinatorial optimisation problems continues to motivate the search for fundamentally new computational paradigms \cite{FarhiSipser2000, FarhiPreda2001,GloverDu2022}. 

Quantum optimisation offers a novel paradigm of solving certain problem classes through e.g. quantum annealing \cite{KadowakiNishimori1998,DasChakrabarti2008,HaukeOliver2020,RajakChakrabarti2023}, adiabatic evolution \cite{FarhiSipser2000,FarhiPreda2001}, or quantum-classical hybrid approaches \cite{CerezoColes2021,AstrakhantsevCarleo2023} such as the Variational Quantum Eigensolver (VQE) \cite{PeruzzoOBrien2014,KandalaGambetta2017,HempelRoos2018,KokailZoller2019} or Quantum Approximate optimisation Algorithms (QAOA) \cite{FarhiGutmann2014,WeckerTroyer2015,WurtzLove2021,FarhiZhou2022,BlekosSummer2024}. Problems that efficiently map onto Ising spin models are particularly suitable for these quantum optimisation methods, with examples including the maximum (weighted) independent set (MIS/MWIS) and quadratic unconstrained binary optimisation (QUBO) problems \cite{lucas2014ising}. In all these approaches the solution to the problem is encoded into the ground state of quantum many-body Hamiltonian \cite{ZhangKamenev2024, PerezArmasDeleplanque2024}.

\begin{figure}[t!]
    \includegraphics[width=\columnwidth]{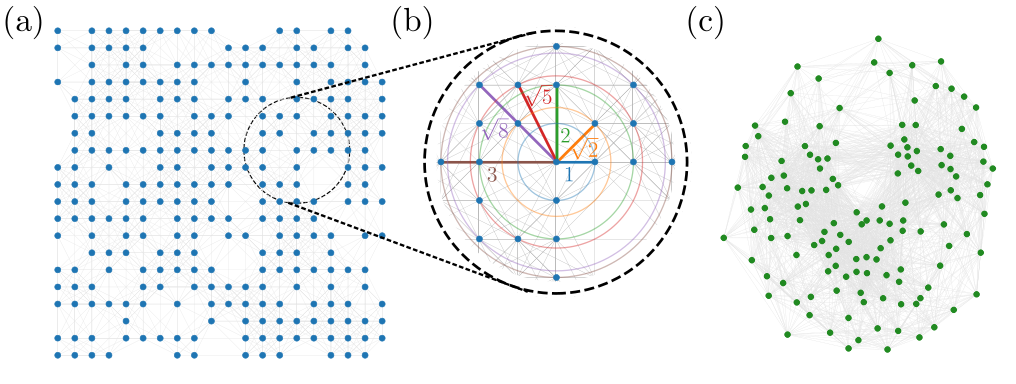}
    \caption{Algorithmic pipeline of the \LAR framework for graphs native to the Rydberg array system. (a) An initial $L\times L$ grid of atoms is randomly populated to density $\rho$, after which native UDG edges are added as determined by the blockade radius $\Rb$. The example shows $L=20$, $\rho=0.7$, $\Rb = 3$. (b) The local vicinity of a node in the graph is shown along with a range of blockade radii, highlight the emerging UDG connectivity. Scaling the blockade radius from $\Rb=1$ to $3$ (see in different colours and associated blockaded circles) increases the maximum number of edge-connected neighbours from 4 to 28. (c) Each UDG graph instance is passed to \LAR which applies classical reductions to create an output kernel with fewer vertices. The kernels are no longer guaranteed to have unit-disk character, and tend to be very dense as demonstrated for the example graph from (a).}
    \label{fig:reduction_pipeline}
\end{figure}

Quantum annealing of Ising spin-models has been demonstrated using superconducting circuit systems which offer tunable couplings and fast annealing timescales, suitable for solving QUBO problems \cite{JohnsonRose2011,desantis26} or quantum simulation \cite{king25}. Open challenges remain around the optimal choice of coupling graphs \cite{ZaribafiyanChangizRezaei2017,pelofske25}, embedding strategies \cite{Choi2008,Choi2011,date19,ceselli23} and scalability in achieving robust demonstrations of quantum advantage for different classes of optimisation problems \cite{kim25,quinton25}.

Among competing architectures, neutral-atom quantum processors based on Rydberg atom arrays have recently emerged as promising candidates for combinatorial optimisation tasks \cite{henriet2020,morgado2021}, combining large-scale, programmable qubit registers with tunable long-range interactions, dynamical reconfigurability \cite{bluvstein22,evered23,bluvstein24,manetsch24,chiu25}, and local addressability \cite{deOliveiraPritchard2025}. Operating in the Rydberg blockade regime where any pair of atoms separated by less than the blockade radius $R_{b}$ cannot be simultaneously excited to the Rydberg manifold \cite{lukin01,LevineLukin2019,saffman10}, these platforms naturally enforce the independence constraint of certain graph optimisation problems, most notably for MIS problems on a restricted class of geometric graphs known as unit disk graphs (UDGs) \cite{pichler2018MIS,pichler2018complexity,ebadi22,kim2024}. Furthermore, local control fields \cite{goswami2024} realising local light shifts, permit embeddings of Maximum Weighted Independent Set (MWIS) problems defined on UDGs \cite{deOliveiraPritchard2025,deOliveiraPritchard2026}. 

\jdp{A fundamental challenge of this approach is that Rydberg interactions generate only UDG connectivity, constrained by the two-dimensional geometry of the planar atomic array, while real-world optimisation problems typically correspond to non-UDG graphs with arbitrary topologies. This has motivated a wide range of encoding strategies for embedding arbitrary problems onto UDG graphs with nearest-neighbour or next-nearest neighbour coupling \cite{kim2022,byun2022,dalyac2023,byun2024QUBO,byun2024HUBO,jeong2023SAT,lanthaler2024,nguyen2023,lanthaler2023,BombieriPichler2024}. However, while these schemes differ in physical details, they share a common limitation: embedding graphs with $N$ logical variables generally results in embeddings with $\mathcal{O}(N^{2})$ physical atoms \cite{nguyen2023,park2024}, and despite recent attempts to further optimise embedded layouts \cite{schuetz2024}, this quadratic overhead remains a significant obstacle for implementing large-scale combinatorial optimisation on near-term neutral-atom platforms.}

It is therefore interesting and timely to investigate natively embeddable graphs more closely. \jdp{To date, most large-scale experimental studies of MIS on Rydberg platforms with several hundred atoms have focussed on native unit-disk instances on a King's lattice (a square lattice with edge couplings to nearest neighbours along horizontal and diagonal paths) \cite{ebadi22,kim2024}. A subsequent analysis however demonstrated that King's lattice UDG MIS instances can be solved to optimality at much larger scales (reaching thousands of nodes) by using classical algorithms on commodity computing hardware, and without relying on instance-specific fine-tuning \cite{andrist2023}.} 

\jdp{Recent works extending these findings introduce hybrid classical–quantum workflows combining graph reduction, embedding, and compilation for Rydberg-based optimisation \cite{schuetz2024,SchuetzKatzgraber2025}. Graph reduction is a classical pre-processing step that acts to simplify a target problem graph by sequential application of reduction rules to create a subset of smaller graph kernels which can be solved more efficiently than direct solution of the initial graph. Highly reducible graphs are deemed `easy' to solve, and in these works the King’s lattice MIS instances were identified as being as algorithmically easy and efficiently reduced from thousands of nodes to smaller kernels with tens or hundreds of nodes. However, only a limited set of four basic reduction rules were employed in these studies, leaving open the question of how this `easy–hard' landscape might shift under state-of-the-art, graph-theoretic kernelisation techniques \cite{LammWerneck2017, HespeStrash2019, LammZhang2019}.}

In this work, we revisit and extend these studies with the aim of identifying the new boundary between `easy' and `hard' instances for native Rydberg graphs, with the workflow schematically depicted in \figref{fig:reduction_pipeline}. Specifically, we investigate whether modern, more powerful reduction pipelines can render previously `hard' instances tractable, and how vertex weights (MWIS) and longer-range native connectivities (corresponding to larger blockade radii) affect problem hardness. Rather than addressing the embedding challenge directly, we here focus on the native 2D UDG problem class, assuming either directly realizable graphs or ones that have already undergone embedding using existing schemes discussed above. We utilise a comprehensive open-source reduction framework \cite{GrossmannSchulz2024} featuring over twenty reduction rules of varying complexity, and systematically quantify the reduction power and hardness scaling across MIS and MWIS instances for random Rydberg arrays of varying size and connectivity, as set by the blockade radius. By interpreting the reducibility as a quantitative measure of instance hardness, we establish a refined understanding of which native Rydberg problems are classically `easy', and crucially which remain intrinsically challenging — thereby delineating the most promising domains to focus further research efforts and for potential quantum advantage. We emphasise that this is not the same definition of problem hardness as the worst-case notions of complexity-theory, but rather pragmatically motivated by and relevant for the context of current near-term quantum devices.

The remainder of this paper is organized as follows. In \secref{sec:background}, we review the background of the MIS and MWIS problems, and summarise the main graph reduction and kernelisation techniques relevant to this study. \secref{sec:results} presents our results and discussion, analysing reduction performance, hardness trends, and the effects of weights and blockade radius on instance complexity. Finally, we conclude in \secref{sec:conclusions} with a summary of our findings, and an outlook on future directions.

\section{Background}
\label{sec:background}

\subsection{MIS/MWIS}
\label{subsec:mwis_background}
The maximum independent set (MIS) problem is a paradigmatic task in combinatorial optimisation. Given a graph $G = (V, E)$  comprised of a set of vertices $V$, and a collection of edges $E$ between pairs of vertices, an \emph{independent set} is a subset $S \subseteq V$ such that no two vertices in $S$ share an edge. The MIS problem seeks an independent set of maximum cardinality. More generally, in the maximum weighted independent set (MWIS) problem, each vertex $i \in V$ carries a weight $w_{i}$, and the objective is to find an independent set $S$ maximising the weight $\sum_{i \in S} w_{i}$.

Both MIS and MWIS can be formulated as a classical minimisation over binary variables $x_{i} \in \{0,1\}$, where $x_{i} = 1$ indicates that vertex $i$ is included in the candidate solution. The classical cost function is given by
\begin{equation}
C(\mathbf{x}) = -\sum_{i \in V} w_{i} x_{i} + U \sum_{(i,j)\in E} x_{i} x_{j} ~,
\end{equation}
where the first term rewards selecting high, positive-weight vertices, while the second assigns a penalty $U > \max_{i} w_{i} > 0$ for any adjacent pair enforcing the independence constraint. For the unweighted MIS problem $w_{i} = 1$, and $U > 1$. Finding the optimal assignment $\mathbf{x}^{*} = \min_{\mathbf{x}} C(\mathbf{x})$
is NP-hard in general, even for the unweighted case \cite{Karp1972,GareyJohnson1978,JohnsonGarey1979,lucas2014ising}, and can be natively realised on UDGs using Rydberg atom tweezer arrays \cite{pichler2018MIS,ebadi22}. Consequently, MIS and MWIS serve as canonical benchmarks for evaluating the performance of both classical and emerging neutral-atom quantum optimisation approaches.

\subsection{Classical Graph Reductions for MIS and MWIS}
\label{subsec:reduction_background}
State-of-the-art classical approaches to the maximum (weighted) independent set (MIS/MWIS) problem make extensive use of \emph{data reduction} or \emph{kernelisation} rules, which systematically reduce a graph instance while provably preserving optimality. Many reduction rules have been developed for MIS and MWIS problems, and are employed in state-of-the-art algorithms~\cite{AkibaIwata2016,HespeStrash2020,SzaboZavalnij2019,PlachettavanderGrinten2021,HespeSchorr2021,LangedalSanders2024,LammZhang2019,WarrenHicks2006,AndradeWerneck2012,GellnerZavalnij2021,EbeneggerdeWerra1984,AlexedeWerra2003,ZhengYu2020,XiaoChen2024,FigielNiedermeier2022,FangXu2016,HeldSewell2012,JiangManya2017,ButenkoStetsyuk2002,ButenkoTrukhanov2007,Strash2016,ChangZhang2017,ButenkoStetsyuk2009,HespeStrash2019,LammWerneck2017,GrossmannStrash2023,GrossmannSchulz2024} 
as a preprocessing step before running an exact and heuristic solver on the reduced graph, and subsequently lifted the optimal solution for the original graph from the solution of the reduced kernel.

\subsubsection{Reduction Philosophy}
Given a graph $G=(V,E)$ with \jdp{$n_{G}$ vertices and} vertex weights $w:V\rightarrow\mathbb{R}_{\geq 0}$, a reduction rule identifies a set of vertices whose membership in an optimal solution is fully determined by its local structure. Removing these vertices and adjusting the objective value (in the weighted case) yields a smaller problem instance $K=(V',E')$ \jdp{with $n_K$ vertices}, termed the \emph{kernel}. Applying all admissible reductions until convergence can often reduce real-world instances dramatically, and the kernel can then be solved by exact branching, branch-and-reduce, or local-search strategies, after which the removed parts of the solution are reconstructed. 

\jdp{
We introduce the reducibility $\xi = 1 - n_{K} / n_{G}$, which serves as a practical proxy for problem hardness, where $n_{K/G}$ denotes the number of vertices in the original ($G$) and reduced kernel ($K$) graphs respectively. Highly reducible graphs ($\xi \approx 1$) are deemed `easy' to solve, whereas graphs with $\xi \gtrsim 0$ remain `hard'. }

\subsubsection{Learned Reduction: KaMIS's \LAR}
\jdp{In Appendix~\ref{sec:reductionrules} we provide examples of some of the rules relevant for MIS and MWIS reductions.} Although many reduction rules are inexpensive, several of the most powerful ones - such as advanced neighbourhood-dominance checks, generalised twin rules, or deep folding cascades - can be costly to evaluate. The \emph{LearnAndReduce} framework \cite{GrossmannSchulz2024} incorporates machine learning to accelerate these expensive reductions. The approach trains a graph neural network (GNN) to predict the likelihood that a reduction rule will apply at a vertex. During kernelisation, for each vertex $v$ the GNN outputs a probability $p(v)$ indicating how promising $v$ is for a computationally heavy rule.  The solver then evaluates those costly reductions \emph{only} on vertices with high $p(v)$, effectively steering the kernelisation process. After reductions saturate, the resulting kernel is passed to a high-performance parallel local-search algorithm (e.g. CHILS - Concurrent Hybrid Iterated Local Search \cite{GrossmannSchulz2024, GrossmannSchulz2025}) integrated within KaMIS. This hybrid of classical kernelisation with learned heuristics constitutes one of the strongest state-of-the-art classical frameworks for large MIS and MWIS instances, enabling reductions that would otherwise be too computationally expensive to deploy at scale.

\section{Results and Discussion}
\label{sec:results}
In the following we present detailed results of numerical experiments employing \LAR to solve randomly generated UDG instances. We systematically vary the underlying array size $L$ and graph connectivity by tuning the blockade radius $R_{b}$ and analyse the reduction statistics of the remaining finite kernels, see \figref{fig:reduction_pipeline} for a schematic illustration. Our work considers blockade radii up to $\Rb = 3$ (throughout this manuscript we work in units of the natural lattice spacing, and set $a = 1$), initially exploring the impact of the improved reduction algorithms on MIS problems whilst comparing to earlier works with simplified reduction rules \cite{andrist2023,schuetz2024,SchuetzKatzgraber2025}. Crucially, we then extend the scope of our study by looking at weighted MWIS problems.

\subsection{MIS}
\label{subsec:mis_results}
We begin by studying native, randomly generated MIS problem instances on square arrays of varying size $L \in \{10,20,30,40,50\}$ and connectivity $1 \leq R_{b} \leq 3$. The generated UDG instances span a wide range of connectivity patterns with the neighbourhood of a generic node graph with node density $\rho = 1$ varying in size between $4$ and $28$ respectively. 

\begin{figure}[t!]
    \includegraphics[width=\columnwidth]{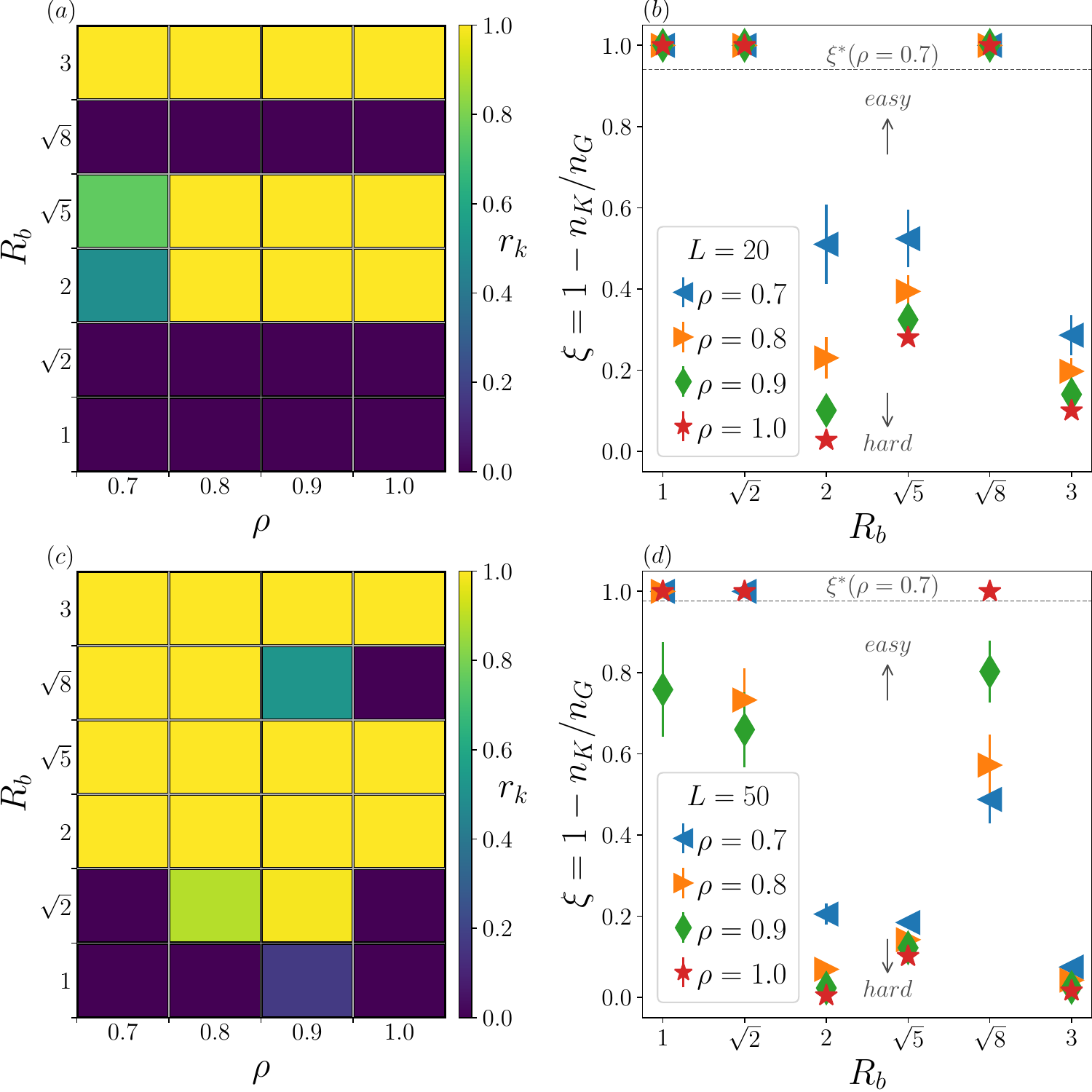}
    \caption{(a) Results for the $L=20$ MIS problems, and (b) their corresponding (mean) reduction factors with errorbars denoting the standard deviation. (c) The bottom row shows the same results as the top for the largest considered array of size $L=50$, with corresponding reduction factors shown in (d). Note that for $\rho=1.0$ there is only a single geometric node arrangement ($\# G = 1$) but for all other parameters, $\rho < 1.0$,  we generated $\# G = 1000$ random graph realisations for our statistical ensemble. The numbers in the left subplots indicate the number of finite kernels $\# K$ from the ensemble. The dashed line in right subplots marks $\xi^{*} = 1 - 1 / \sqrt{N}$, the threshold above which it is more resource efficient to embed the reduced kernel - here chosen for $\rho = 0.7$ (higher densities would result in an even larger threshold $\xi^{*}$).}
    \label{fig:MIS}
\end{figure}

\figref{fig:MIS} (a) shows an example of a $L=20$ square array across different node densities $\rho \in \{0.7, 0.8, 0.9, 1.0\}$. For each problem instance $(L, \rho, \Rb)$ we generate $\# G = 1000$ random graph instances (except for $\rho=1.0$, where only $1$ exists). We record the number of finite kernels ($\# K$), shown in each box of \figref{fig:MIS} (a) and (c), the corresponding reduction factors $\xi$, and plot the kernel ratio $r_{k} = \# K / \# G$ as a heatmap. We observe a sharp transition in the number of finite kernels as a function of blockade radius across all densities shown here. For $\Rb \leq \sqrt{2}$ (nearest, and next-nearest neighbour connectivity) every single randomly generated graph instance was solved to optimality resulting in an empty kernel ($\xi = 1$), indicating that graphs at this size, density and connectivity do \emph{not} pose a challenge for existing state or the art classical reduction techniques. We have also confirmed that \LAR was able to fully solve less dense graphs ($\rho < 0.7$) generated in this fashion across all studied connectivities (not shown here). Conversely, for high densities $\rho \geq 0.7$ (with the exception of $\Rb = \sqrt{8}$), all other generated problem instances ($\Rb \geq 2$) retained a finite kernel, demonstrating an easy-hard transition from the perspective of the \LAR algorithm. A blockade radius of $\Rb = \sqrt{8}$ appears to be the outlier and is surprisingly easy to be reduced by standard techniques. This observation is in agreement with a time-to-solution (TTS) analysis of an exact B\&B (branch and bound) solver for native UDG graphs \cite{andrist2023}.

To better understand the structure of these graphs, \figref{fig:MIS} (b) shows the distribution of the reduction factor $\xi$ across the \emph{finite} kernels shown in (a). Excluding the fully reduced graphs for $\Rb \leq \sqrt{2}$ and $\Rb = \sqrt{8}$, we note that reduction factors are generally relatively low, $\xi < 0.5$, indicating these graphs are already not trivial. 

In particular we observe that the kernelisation techniques show a pronounced peak in $\xi$ for the intermediate regime $2 < \Rb < 3$, but appear to struggle for blockade radii of $2$ and $3$. In \cite{andrist2023} the TTS for an exact sweeping line algorithm (SLA), and a commercial B\&B solver was examined as the graph connectivity was tuned. It was found that problem hardness (as measured by the TTS) can be systematically tuned over several orders of magnitude, with the correct behaviour in the asymptotic cases of $0$-regular ($\Rb \rightarrow 0$) and complete ($\Rb \rightarrow \sqrt{2}L$) graphs, as well as pronounced peaks in the TTS at integer blockade radii of $\Rb = \{2,3,4\}$. Here we corroborate that $\Rb = 2,3$ are not only challenging connectivities for exact solvers, but also for classical reduction algorithms, while $\Rb = \{\sqrt{5}, \sqrt{8}\}$ present `easier' problem instances for these classical techniques, with similar behaviour across all densities shown here, and an overall trend towards lower $\xi$ values as $\rho$ increases.

In line with the findings in \cite{andrist2023} we find that dense, large graphs ($\rho \geq 0.7$ and $L\gtrsim 40$) become less tractable, and result in kernels with much lower reduction factors, even for the King's lattice connectivity at $\Rb = \sqrt{2}$ and $\Rb = \sqrt{8}$. \figref{fig:MIS} (c) shows the same analysis for the largest studied array of $L=50$. The vast majority of graphs did yield a finite kernel with e.g. the largest mean reduction factor for $\rho=0.9$ (at $\Rb = \sqrt{5}$) dropping by a factor of $3$ from $\bar{\xi} \sim 0.35$ ($L=20$) down to $\bar{\xi} \sim 0.12$ for $L=50$. The \jdp{other blockade ranges show} even smaller reductions, and we again find very similar behaviour for the different fillings $\rho$ in the array. Even for $\Rb = \sqrt{8}$ and $\rho < 1$ did we find graphs which are not fully reducible, where arrays with a \jdp{lower} density appear to be harder from a kernelisation perspective.

Finally, as noted above, using the full reduction suite of \LAR (in particular the most powerful techniques) means that the resulting graph kernels are no longer guaranteed to be UDG native. This brings us back to the embedding problem outlined at the beginning of this article. Current embedding schemes still require a quadratic resource overhead for embedding non-UDG instances into UDG arrays. This implies that a kernel with a reduction factor $\xi$ requires $\sim (1 - \xi)^{2}N^{2}$ nodes to be embedded natively on a neutral atom platform. From a pure resource perspective this leads us to conclude that there is a natural trade-off between embedding a reduced kernel, and solving the original native graph directly, with the former only being favourable for large reductions $\xi \gtrsim \xi^{*} = 1 - 1 / \sqrt{N}$. For a dense completely filled ($\rho = 1$) graph of modest size $L=10$, this already suggests that we would require reductions in excess of $90\%$ before embedding the resulting kernels becomes beneficial. None of the investigated kernels have exhibited such large reductions, which leads us to identify larger, more connected problem instances as target problem areas where quantum algorithms can be useful, and which may lead to a potential quantum speedup for future experiments with Rydberg atom arrays.

\subsection{MWIS}
\label{subsec:mwis_results}
We now turn to the maximum \emph{weighted} independent set (MWIS) problem on native, randomly generated UDG instances. We here focus on the largest array size $L=50$, as its unweighted MIS problem presented the richest finite kernel landscape. We also confirmed that we obtain qualitatively very similar results for the smaller graphs. For each geometric instance ($L$, $\rho$ and $\Rb$) we explore the impact weighting has on the reduction distribution by assigning random integer vertex weights sampled from a uniform distribution $w_{i} \in [1,W]$, where $W\in\{10,100,1000\}$. For each triple $(\rho, \Rb, W)$ we randomly select $100$ different geometries (from the previous MIS data) and, for every such geometric graph instance produce $100$ independent weight realisations. As before, we record the resulting reduction factors $\xi$. We have studied the effect the \emph{range} of the weighting might have on the reduction factors, but found qualitatively very similar results for all cases \jdp{(see App.~\ref{sec:w} for details)}. In the following, unless otherwise stated, we present results for $W = 10$.

\begin{figure}[t!]
    \includegraphics[width=\columnwidth]{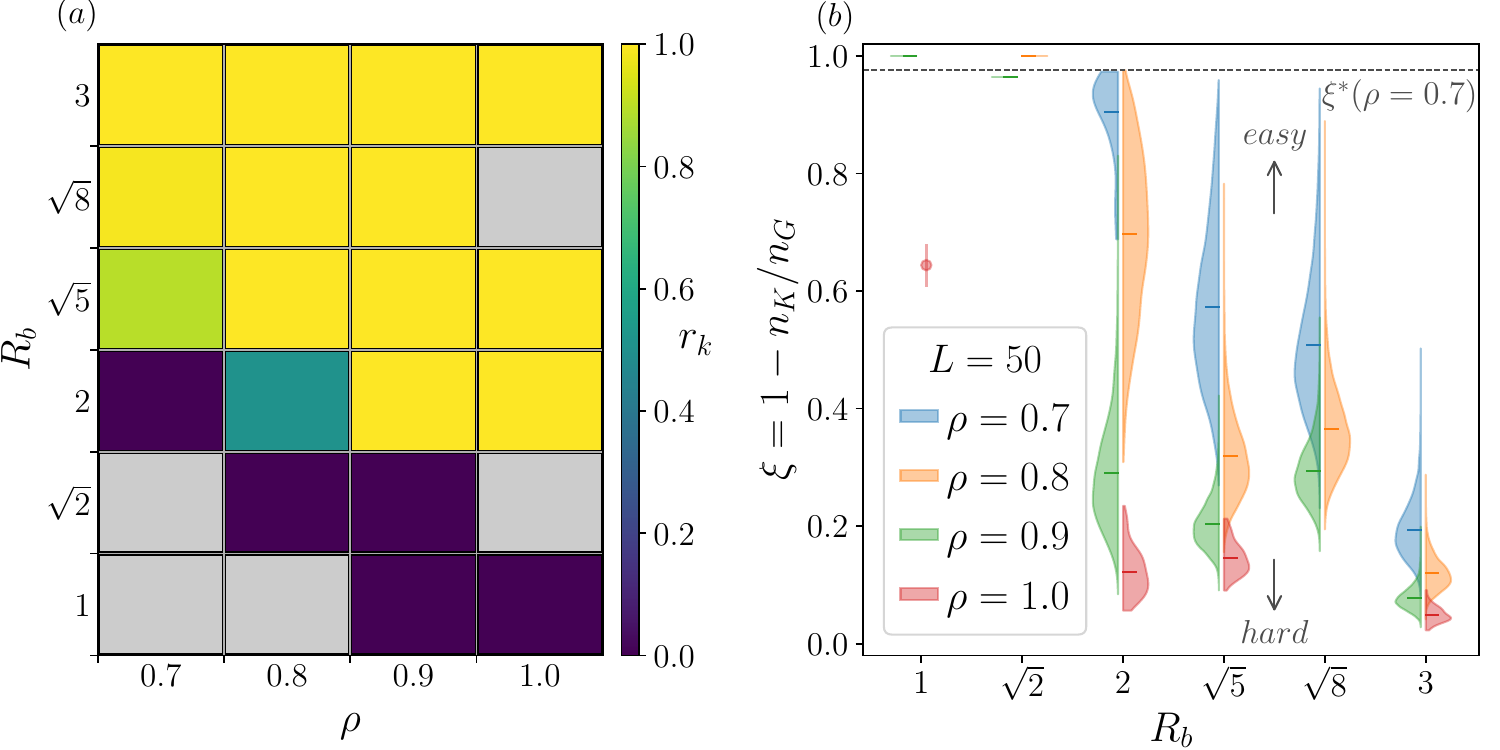}
    \caption{\jdp{Reducibility of native MWIS graphs. (a) Results for $L=50$, where data are generated by randomly sampling 100 graph geometries for each combination of $(\rho,R_b)$ (one for $\rho=1$), and for each then generating a further 100 unique vertex weightings using uniform random weights with a range of $W=10$. For certain parameters the unweighted MIS problem was fully reducible, in which case we did not run any weighted problems and highlighted this with a grey, blank square. (b) Reduction factor distribution as half-violins for all node densities $\rho$ and blockade radii $\Rb$. The dashed line marks $\xi^{*} = 1 - 1 / \sqrt{N}$, the threshold above which it is more resource efficient to embed the reduced kernel - here chosen for $\rho = 0.7$ (higher densities would result in an even larger threshold $\xi^{*}$).}}
    \label{fig:MWIS}
\end{figure}

\figref{fig:MWIS} (a) illustrates typical UDG-MWIS instances for $L=50$ as a function of blockade radius and node density. A first notable observation is that adding weights can make the instances \emph{more} amenable to classical reductions: whilst every single unweighted problem instance for $\Rb \geq 2$ (with the exception of $\Rb = \sqrt{8}$) was deemed `hard' (i.e. had a finite kernel), here we see that certain weighted instances are solved to optimality (empty kernel). Furthermore, \figref{fig:MWIS} (b) shows that we find larger mean reduction factors compared to the MIS case. This behaviour is consistent with the interpretation that adding weights can partially lift the degeneracies present in the MIS problem, thereby providing the reduction machinery with more structure to exploit. 

These conclusions are further corroborated by known approximability theory results for the MIS and MWIS problem; UD-MWIS is in the polynomial-time approximation scheme (PTAS) class \cite{ErlebachSeidel2001, Matsui2000, DasSurKolay2015, NandyRoy2017, BonamyThomasse2018}, while the general MIS problem is APX-complete \cite{GareyJohnson1978, Halstad1999, BazganPaschos2005}. Thus MIS is believed not to admit a PTAS algorithm, and cannot be in the PTAS class (unless P = NP) \cite{deCorrecBertrand2025}. Likewise following the quantum PCP conjecture \cite{AharonovVidick2013}, MIS is believed to remain in the quantum equivalent of the APX class even with quantum algorithms.

It should be noted that weighting in turn could lead to \emph{smaller} spectral gaps generically for the logical problems, which would on the other hand lead to adiabatic bottlenecks in quantum annealing procedures. Nevertheless, even in the weighted setting we still observe predominantly modest reduction factors, with $\xi \lesssim 0.5$ for many instances, indicating that these problems remain non-trivial in the majority of cases, and only very few outliers come close to the reduction target $\xi^{*}$ above which embedding of the reduced kernel might be favourable over solving the original native graph instance. 

To quantify when MWIS remains challenging, \figref{fig:MWIS} (b) shows the $\xi$ distribution for the finite kernels across the different blockade radii and for all node densities $\rho$. Generally we find that weight-induced simplifications shift the distribution towards higher values of $\xi$ as compared to the unweighted MIS data in \figref{fig:MIS}. In particular graphs with a relatively local structure (e.g. $\Rb < 2$) are solved to near optimality even for graphs with few defects ($\rho = 0.8$ or $0.9$). In contrast, for the longer-range connectivity $\Rb = \{\sqrt{5}, 3\}$, we find that kernelisation performance sharply deteriorates: the distribution becomes more skewed towards lower values of $\xi$, exhibiting a long tail corresponding to the minority of instances where reductions are more sizeable ($\xi$ larger than its mean $\bar{\xi}$). \jdp{For $\Rb=\sqrt{8}$, which was efficiently reducible for small MIS instances, we observe that weighting acts to reduce the mean $\xi$ relative to MIS (e.g. for $\rho=0.9$ the mean drops from 0.85 to 0.3) whilst still proving more reducible than instances with $\Rb = \{\sqrt{5}, 3\}$.} The overwhelming majority of weighted graphs therefore remain difficult to reduce. This mirrors the behaviour seen in MIS, suggesting that here too longer-range UDG structures are intrinsically more resistant to classical reductions regardless of the weight scale. Since weights and geometry both influence reducibility, it is natural to ask whether MWIS hardness is primarily set by the underlying connectivity pattern or by the choice of weights. While a full geometry-resolved study is beyond the scope of this work, we note that for a fixed geometry the spread of reduction factors across different weight realisations can be substantial, particularly for moderate blockade radii. This indicates that both sources contribute in a non-trivial way: geometry determines a baseline difficulty, whereas weights modulate how effectively reduction rules can break symmetries or eliminate ambiguous local configurations.

\begin{figure}[t!]
    \includegraphics[width=\columnwidth]{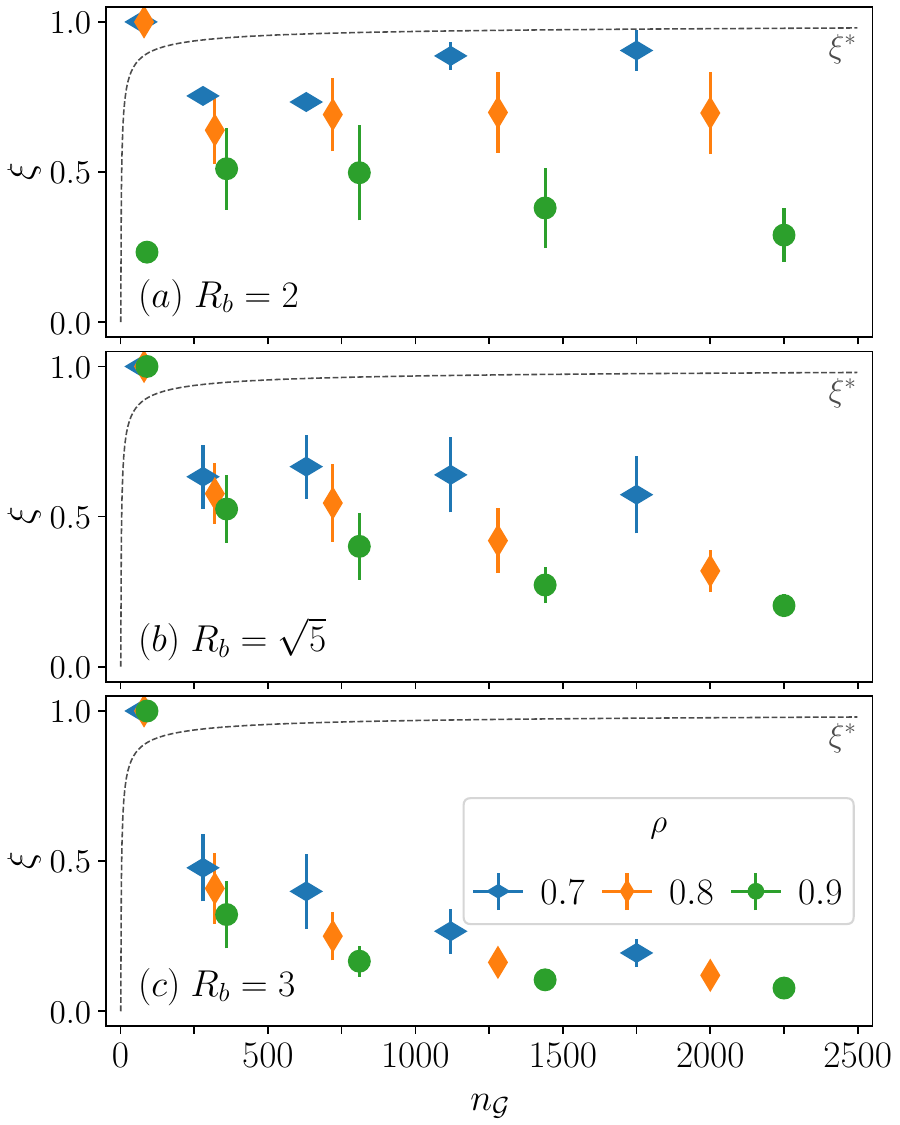}
    \caption{Scaling of the reduction factor $\xi$ with the graph size $n_{G} = \vert G \vert = \rho L^{2}$ for three different node densities $\rho$, and connectivities $\Rb$ (a-c) as depicted in the legend. The grey dashed line indicates the critical reduction $\xi^{*}$ beyond which an embedding of the kernel is preferable (from an atom number resource perspective) compared to solving the original graph.}
    \label{fig:MWIS_ReductionScaling}
\end{figure}

We conclude this section by studying how the reduction factors behave as a function of the size $n_{G} = |G|$ of the graphs, shown in \figref{fig:MWIS_ReductionScaling}. At moderate connectivity ($\Rb = 2$) $\xi$ displays a very curious behaviour for the different node densities. 
For $\rho = 0.7$ the reduction factor dips to a minimum for $20 \leq L \leq 30$ but remains very high for very small or large graphs. Slightly denser graphs do not exhibit the resurgence of $\xi$ for large arrays but plateau to a steady value of $\xi \sim 0.7$. Increasing the node density further inverts the observed behaviour completely. The smallest $L=10$ graph has the smallest reduction factor, which dramatically increases when the array size increases to $20$, after which $\xi$ gradually decreases with system size as one would naively expect. Higher connectivity graphs $\Rb = \sqrt{5}$ and $\Rb = \sqrt{8}$ follow this more consistently. Smaller graphs exhibit higher reductions, which then gradually decrease with increasing system size. Further we can nicely observe that higher density graphs present more of a challenge to these reduction techniques since e.g. $\rho = 0.7$ (blue diamonds) reduction factors are consistently larger than higher values of $\rho$.

Finally, as in the MIS case, invoking the full reduction suite of \LAR means that resulting MWIS kernels are generally no longer UDG-native. For $L = 50$, this corresponds to an even larger reduction target of $\xi^{*} \sim 0.98$. Across all MWIS instances we examined - excluding the fully reducible cases - no kernels achieved such dramatic reductions. This leads us to the same conclusions as in the unweighted MIS case: in most practical situations it may preferable to run the \emph{original} native MWIS instance on a Rydberg atom processor rather than embedding a reduced kernel. Consequently, large, dense, and longer-range UDG-MWIS instances emerge as further candidates for probing the computational frontier of near-term neutral-atom hardware. 

\jdp{We note that for the discussed results we have used \LAR with the GNN filter enabled. However, we find no change in conclusion when implementing the complete set of reduction methods. This confirms that a lack of reducibility is not a consequence of the GNN filter having been trained on an unsuitable class of graphs, but instead is due to native, high-connectivity graphs being intrinsically harder to reduce (see Appendix~\ref{sec:GNN} for details). }

\section{Conclusion}
\label{sec:conclusions}
In this work we have examined the extent to which state-of-the-art classical reduction techniques can simplify native UDG MIS and MWIS problem instances relevant to Rydberg-atom quantum hardware. Our analysis reveals several clear trends. 

Firstly, even for moderate array sizes, many dense UDG-MIS instances remain \emph{not} fully reducible: sizeable irreducible kernels persist despite the extensive suite of reductions employed. Introducing vertex weights, however, tends to increase reducibility, with some instances that admit finite MIS kernels becoming fully reducible once weights are added, indicating that weights can partially lift degeneracies that otherwise obscure the action of reduction rules.

Secondly, non-trivial kernels arise predominantly in moderately large and dense systems, typically for $L \gtrsim 20$ and $\rho \gtrsim 0.8$. Within this regime, the structure of the underlying UDG connectivity plays a decisive role. Longer-range connectivities - corresponding to larger blockade radii - systematically degrade the performance of classical preprocessing, leading to broad distributions of low reduction factors and signalling that these graphs remain intrinsically difficult for kernelisation.

For instances that do retain finite kernels, executing these kernels on a Rydberg quantum processor necessitates embedding them into a native UDG layout. Current embedding schemes incur a quadratic resource overhead, implying that only kernels with reduction factors $\xi \gtrsim 1 - 1/\sqrt{N}$ are advantageous from a purely resource-based perspective. Even for a modest $N=100$ node graph this would require $\xi > 0.9$, a level of reduction essentially absent from our data (excluding the fully reducible cases). This indicates that, in the overwhelming majority of practically relevant situations, direct execution of the \emph{original} native graph may be preferable to embedding a reduced kernel.

Taken together, these results identify a clear region of parameter space - dense, moderately large arrays with extended UDG connectivity - which provides a natural target for assessing the computational capabilities of near-term neutral-atom quantum processors. \jdp{To connect these parameter regimes to real systems, typical platforms offer atom spacing around $a=4~\mu$m, and two-photon Rabi frequencies $\Omega/2\pi\sim2$~MHz \cite{aquila}. The blockade radius is given by $\Rb=\sqrt[6]{\vert C_6\vert/\Omega}$, where is $C_6$ is the dispersion coefficient for the Rydberg interaction strength which scales as $n^{11}$ for $nS_{1/2}$ Rydberg states. To achieve $\Rb>3a$ requires $\vert C_6\vert>6\times10^6$~MHz$\cdot\mu$m$^6$, corresponding to choosing $n\ge83$ in Rubidium making this high-connectivity regime experimentally feasible \cite{sibalic17arc}.}

Our findings open several compelling avenues for follow-up work. An important next step is to determine how exact classical solvers (ideally in a co-designed pipeline together with existing reduction algorithms) perform on these native graphs, and to pinpoint when they begin to struggle. Likewise, it remains to be tested whether quantum annealing or adiabatic strategies remain viable at these problem sizes, particularly beyond King's lattice connectivity. If an embedded finite kernel becomes a weighted King's lattice instance, and if such MWIS problems indeed remain tractable for classical methods, then embedding - despite its overhead - could become competitive in certain regimes. \jdp{A detailed comparison of time-to-solution across reductions, exact solvers, and quantum hardware will be essential for establishing which classes of UDG-MIS and UDG-MWIS problems constitute meaningful benchmarks for future experimental demonstrations. Finally, to exploit the intrinsic advantage offered by the higher native connectivity it is essential to develop new embedding strategies that can map real-world graph problems onto MIS or MWIS instances with $R_b>\sqrt{2}$.}

\begin{acknowledgements}
We thank Ernestine Großmann, Kenneth Langedal, Arthur La Rooij, Gerard Pelegrí, Paul Warburton, Martin J. A. Schuetz, Ruben S. Andrist, and Helmut G. Katzgraber for useful discussions. Particular thanks to Ernestine Großmann and Kenneth Langedal for their help in setting up and running \LAR. This work was supported by EPSRC through grant numbers EP/Z53318X/1 and EP/T005386/1. The data presented in this work are available at \cite{kombe26data}.
\end{acknowledgements}

\appendix
\section{Examples of Fundamental Reduction Rules}
\label{sec:reductionrules}
In the following section we provide examples of some of the rules relevant for MIS and MWIS reductions.

\subsubsection{Examples of Fundamental Reduction Rules}
In the following we give a brief description of a subset of reduction rules used in \LAR. A complete list and details can be found in Table $1$ of \cite{GrossmannSchulz2025}.
\paragraph{Isolated Vertex.}
If a vertex $v$ has a open neighbourhood (simply called the neighbourhood) $N(v)= \{ u \mid (v, u) \in E\} = \emptyset$, then $v$ must belong to every MIS/MWIS solution. Thus one includes $v$ in the MIS/MWIS, removes it from $G$ to form $K$, and records a weight offset of $w_{v}$.

\paragraph{Simplicial Vertex.}
A vertex $v$ is \emph{simplicial} when $N(v)$ induces a clique.  
For MWIS, if $w_{v} \ge \max_{u\in N(v)} w_{u}$, then selecting $v$ dominates any choice of its neighbours, and one includes $v$ in the MWIS, removes the closed neighbourhood $N[v] = \{v\}\cup N(v)$, and records a weight offset of $w_{v}$. Otherwise, $v$ can be removed \cite{AkibaIwata2016, LammZhang2019}.

\paragraph{Pendant (Degree One) Vertex.}
If a vertex $v$ has exactly one neighbour $u$, then an optimal solution must include $v$ whenever $w_{v} \ge w_{u}$. In the unweighted case, the pendant vertex $v$ always dominates the neighbour $u$. Accordingly, $v$ is selected, and both $v$ and $u$ are removed. If $w_{v} < w_{u}$, we form $K$ by removing $v$, updating $w_{u} \rightarrow w_{u} - w_{v}$, and recording the weight offset $w_{v}$. It holds that $u \in \text{MWIS}(G)$ iff $u \in \text{MWIS}(K)$ \cite{GuPeng2021}.

\paragraph{Twin Reduction.}
Let $u,v \in V$ have equal, independent neighbourhoods $N(u)=N(v)=\{p,q,r\}$. Any optimal solution contains either both $u,v$ or neither.  
If $w(\{u,v\}) \geq w(\{p,q,r\})$, include $u,v$ and reduce graph to $K = G - N[\{u,v\}]$, with an offset $w_{u} + w_{v}$.  
If $w(\{u,v\}) < w(\{p,q,r\})$ but $w(\{u,v\}) > w(\{p,q,r\}) - \min\{w_{x} \mid x \in \{p,q,r\}\}$, fold $\{u,v,p,q,r\}$ into new vertex $v'$ with $N(v') = N(\{p,q,r\})$, and $w_{v'} = w(\{p,q,r\}) - w(\{u,v\})$. The new graph is given by $K = G((V \cup \{v'\}) \setminus N[\{u,v\}])$ with an offset of $w(\{u,v\})$. To reconstruct the solution for $G$, one selects $\{p,q,r\}$ if $v'$ is in the solution to $K$, and $\{u,v\}$ otherwise. Note, that the twin rule can be generalised to larger neighbourhoods, and ones which are not an independent set \cite{LammZhang2019}.

\paragraph{Unconfined Reduction.}
Let $S \subseteq V$ be an independent set, and assume $S$ is contained in every MWIS of $G$. A vertex $x \in N(S)$ is a child of $S$ if $w_{x} \ge w(S \cap N(x))$. It is an extending child if there exists $y \in N(x)\setminus N[S]$ such that $w_{x} \ge w(S \cap N(x)) + w(I_y)$, where $I_y$ is an MWIS of $G[(N(x)\setminus\{y\}) \setminus N[S]]$. The vertex $y$ is called a satellite, and it can be shown that every MWIS contains the satellites from each extending child $x$ of $S$. Now, iteratively extend $S=\{v\}$ by adding satellites of extending children. If at some step there exists a child $x$ with $w_{x} \ge w(S \cap N(x)) + w(G[N(x)\setminus N[S]])$, then $v$ is unconfined, and can be removed $K = G - v$. If no further satellites exist, $S$ confines $v$, and every MWIS containing $v$, also contains $S$ and one cannot remove $v$ \cite{GrossmannSchulz2025}.

\paragraph{Vertex Folding.}
In the unweighted MIS problem \cite{ChenJia2001}, if a vertex $v$ has exactly two non-adjacent neighbours $u$ and $w$, then every optimal solution includes either $v$ or both $u$ and $w$. Thus $\{u,v,w\}$ can be \emph{folded} into a single super-vertex $v'$. 
For the weighted case, let $N(v)$ be independent. If $w(N(v)) > w(v)$ but $w(N(v)) - \min_{u \in N(v)}\{w_{u}\}< w_{v}$, fold $v$ and $N(v)$ into new vertex $v'$ with $w_{v'} = w(N(v)) - w_{v}$. To reconstruct the MWIS on $G$ one selects $N(v)$ if $v'$ is in the solution to $K$, and $v$ otherwise \cite{LammZhang2019}.

\section{Graph Weighting}
\label{sec:w}
\begin{figure}[t!]
    \includegraphics[width=\columnwidth]{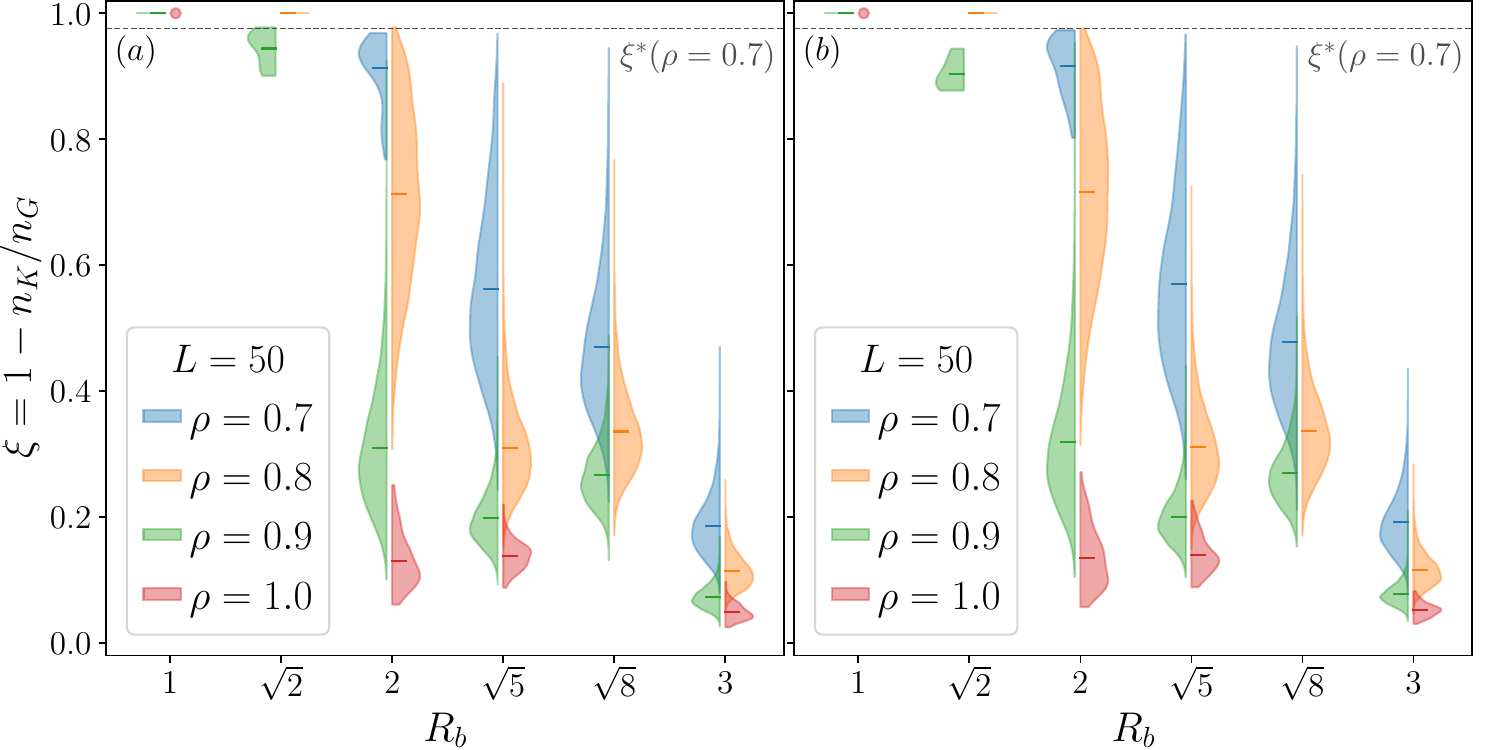}
    \caption{\jk{Comparing the effect of different weight ranges for (a) $W = 100$, and (b) $W = 1000$ on the reductions. The results are do not show any significant differences, suggesting that the size of the weightings $W$ do not affect the overall resulting kernel distribution.}}
    \label{fig:app_MWIS_weightings}
\end{figure}

\jdp{In investigating the efficacy of graph reduction methods on weighted graphs we implement random vertex weightings chosen uniformly from $[1, W]$, with weight ranges $W=\{10, 100, 1000\}$. Figure~\ref{fig:app_MWIS_weightings} shows the results for the reduction of graphs with array size $L=50$, and weight ranges $W=100$ (a) and $W=1000$ (b). Comparing with the data above in Fig.~\ref{fig:MWIS}(b) for $W=10$, we find no significant change in the performance of the \LAR reduction algorithm. We are thus confident that our conclusions above based on weighted data for $W=10$ extend to all graphs with a uniform weighting $W>1$.}

\begin{figure}[t!]
    \includegraphics[width=\columnwidth]{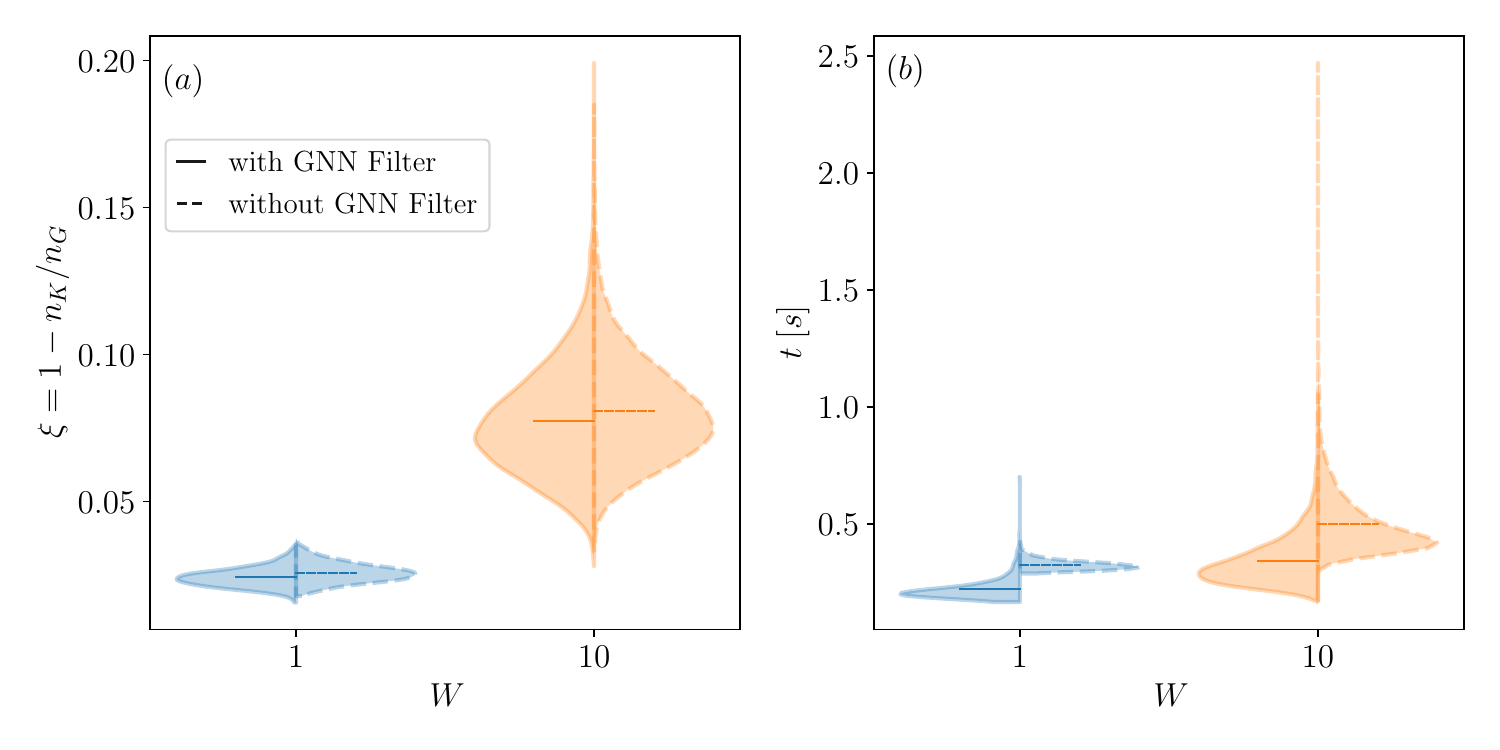}
    \caption{\jk{Comparing the reduction distribution and runtime of the \LAR algorithm when employing the GNN (left half-violin, solid line), or applying the reduction rules exhaustively (right half-violin, dashed line). The overall distribution of reduction values in (a) is shifted to slightly larger values, indicating that employing the full reduction suite exhaustively can bring about smaller kernels on average. The performance gain is however almost negligible, and in reality must be balanced with a noticeable increase in the runtime (excluding I/O timings) as shown in (b). Data shown for $L=50$, $\rho=0.9$ and $R_b=3$ corresponding to some of the least reducible graph examples found above.}}
    \label{fig:app_runtime}
\end{figure}

\section{GNN Filtering and Runtime}
\label{sec:GNN}
\jdp{Whilst the focus of this work is on exploring regimes in which native Rydberg graphs are robust to reduction methods, it is relevant to consider both the impact of the graph neural network (GNN) methods employed by the \LAR algorithm, and its runtime. For the presented results above, \LAR was run with the GNN filter enabled attempting to identify promising nodes and guide further reduction techniques. To verify that the corresponding reduction efficiency is not significantly affected by the training set used for the GNN filter, Fig.~\ref{fig:app_runtime}(a) shows the reduction efficiency with (left half-violin, solid line) and without (right half-violin, dashed line) the GNN Filter enabled for some of the hardest graph configurations we observed ($L=50$, $\rho=0.9$, $R_b=3$, and with weight ranges $W=1$ for MIS and $W=10$ for MWIS instances). The left subplot (a) shows that the kernel size, using the complete set of reductions, is only marginally smaller (i.e. a $\xi$ distribution shifted to slightly larger values) than that achieved using the GNN selected methods. Comparing the runtimes however in Fig.~\ref{fig:app_runtime}(b) shows that the GNN offers an average speedup of around $20\%$, with the runtime for MIS graphs on average around $0.3~s$ for MIS and $0.4~s$ for MWIS configurations with GNN enabled. When considering quantum advantage, time to solution is a key metric. The \LAR method is fast compared to the current runtime of neutral atom platforms with clock rates of a few Hz \cite{aquila,ebadi22}, however the development of continuously loaded platforms \cite{chiu25} and fast readout techniques using single-photon detector arrays \cite{zeng26} provide a pathway to clock-speeds $>1$~kHz.}

\newpage{}

\begin{thebibliography}{112}
\bibitem [{kom()}]{kombe26data}%
  \BibitemOpen
  \bibinfo {title} {{DOI to be added in proof}}\BibitemShut {NoStop}%
\end{thebibliography}
%

\end{document}